\documentclass{DISproc}

\begin{document}
%------------------------------------
\title{Searches for New Physics in the Top Sector at the Tevatron}

%for single authors the superscripts are optional
\author{{\slshape Yvonne Peters$^1$}\\[1ex]
$^1$University of Manchester, Manchester, M13 9PL, England }

% please enter the contribution ID for the DOI
\contribID{xy}

\doi  % if there is an online version we will register DOIs
%FERMILAB-CONF-12-214-E
\maketitle

\begin{abstract}
The top quark, discovered in 1995 by the CDF and D0 collaborations at
the Tevatron collider at Fermilab, is the heaviest known elementary
particle today. Due to its high mass and short lifetime, the top quark plays
a special role in searching for physics beyond the Standard Model.  In
this article, recent results of searches for new physics in the top
sector, performed  by CDF and D0, are presented. In particular, we discuss the
search for $t\bar{t}$ resonances, for $tj$ resonances, the search for
heavy fourth generation quarks, for dark matter produced in
association with single tops, the study of anomalous couplings, the
search for boosted top quarks as well as the analysis of Lorentz
Invariance violation in the top quark sector.
\end{abstract}

\section{Introduction}
Discovered in 1995 by the CDF and D0 collaborations, the top
quark~\cite{CDF_obs, D0_obs} is the heaviest known elementary particle
today, with a mass of $m_t = 173.18 \pm 0.94$~GeV~\cite{topmassworldaverage}. The top quark decays before
hadronization, therefore being  the only particle to study bare
quarks. Furthermore, the Yukawa coupling of the top quark and  the Higgs boson is expected to be
large due to its high mass. The special properties of the top quark
make it an interesting particle to study and as windwo to new
physics.

In the following, recent searches for physics beyond the Standard
Model (SM) in the top quark sector, performed by the CDF and D0
collaborations using Tevatron Run~II data, are
presented.

\section{Searches for New Physics in Top Quark Production}
At the Fermilab Tevatron, a proton-antiproton collider with a center of mass
energy of $\sqrt=1.96$~TeV, top quark production occurs dominantly in
pairs ($t\bar{t}$) through the strong interaction, with about 85\% via $q\bar{q}$
annihilation and about 15\% via gluon-gluon fusion. At about half the
production cross section of $t\bar{t}$, single top quark production
via the electroweak interaction takes place. 

For measurements of the $t\bar{t}$ production cross section and top
quark properties, the
$t\bar{t}$ final states are classified according to the decays of the
two $W$ bosons from the top and anti-top decay. We separate the final states into dileptonic,
semileptonic and allhadronic channels according to the number of leptons in the final state. If the lepton is a hadronic decaying
tau, the events are treated as separate channels ($\tau$+lepton and $\tau$+jets).

In the SM, no $t\bar{t}$ resonances exist, while many models beyond
the SM predict production via a resonance, as for example Topcolor
assisted technicolor models. Using events in the semileptonic final
state, both bthe CDF and D0 collaboration searched for a narrow resonance
$X$, with $\Gamma_X=1.2\% M_X$, by searching for a bump in the spectrum of the invariant
$t\bar{t}$ mass, $m_{t\bar{t}}$. Using events with at least four jets and
4.8~fb$^{-1}$ of data at CDF and at least three jets and 5.4~fb$^{-1}$
  at D0,  limits on
 $\sigma(p\bar{p}\rightarrow X)\times B(X\rightarrow t\bar{t})$ versus
   $M_{X}$ have been extracted. In the benchmark model of topcolor
   assisted technicolor, a $Z^{'}$ for masses below 835~GeV is
   excluded by D0~\cite{reso_d0} and below 900~GeV by
   CDF~\cite{reso_cdf} at the 95\% confidence level (CL). 

Recently, CDF performed a search for a top plus jet ($tj$) resonance $M$ using
the full Run~II data set of 8.7~fb$^{-1}$ by looking for a $tj$
resonance in the $t\bar{t}j$ system. A kinematic fitter is applied on events with at least five jets,
of which at least one has to be identified as a $b$-jet, in
the semileptonic final state, and a bump search in the $tj$ invariant  mass
is performed. Limits are set on  $\sigma(p\bar{p}\rightarrow M
t\bar{t})$,
resulting in upper limits between $0.61$~pb and $0.02$~pb at the 95\%
CL.  These can be translated into limits on  the mass of $M$ assuming
$M$ to be part of a new color singlet or color triplet
model~\cite{topjetreso_cdf}.

%something more on motivation with asymmetry? 

Another search recently performed by CDF using 7.7~fb$^{-1}$
investigates the possibility of a dark matter candidate $D$ produced
in association with a top quark. Single top events, where the top
quark decays fully hadronically and the dark matter candidate leaves
high missing transverse energy in the detector, are used for this
search. A template fit of the missing transverse energy spectrum is
performed in events with at least three jets and no leptons,
inspecting dark matter candidates with masses of up to 150~GeV.
Upper limts on $\sigma(p\bar{p}\rightarrow M
t\bar{t})$ can be set as function of $m_D$~\cite{darkmatter_cdf},
which are about $0.5$~pb over the investigated mass range. 

\section{Searches for New Physics in Top Quark Decay}
In the SM, the top quark decays with a probability of almost 100\%
into a $W$-boson and a $b$-quark. The coupling of the $W$ boson to
fermions has the $V-A$ form of a left-handed vector interaction.
Possible new physics could occur if the coupling of the $W$ boson to
the top and bottom quark ($tWb$ coupling) is of the form of right-handed vector
couplings, or left- or right-handed tensor couplings. In an effective
Lagrangian approach, the different couplings can be introduced as form
factors $f^L_V$, $f^R_V$, $f^L_T$, $f^R_T$, describing the left (L) and
right (R) handed vector (V) and tensor (T) couplings, respectively. In
the SM, $f^L_V=1$ and all others are zero. 

Recently, the D0 collaboration performed a search for anomalous
couplings using information from single top quark production and the
measurement of the $W$ helicity in top quark decays. Using single top
quark events, mutlivariate discriminants are trained on a single top sample with either  $f^R_V$, $f^L_T$, or $f^R_T$ set to one as the
signal sample, while SM single top ($f^L_V=1$) is considered as part
of the background. For each trained multivariate discriminant, the
pair of one of the anomalous couplings form factors and the coupling
form factor $f^L_V$ are
then considered simultaneously, and limits can be extracted in the
plane of    ($f^R_V$,  $f^L_V$) , ($f^L_T$,  $f^L_V$), or ($f^R_T$,
$f^L_V$)~\cite{anom_singletop}.  Furthermore, the $W$ helicity in top
quark decays can be measured using the distribution of the angle
 between the direction opposite to the top
quark and the direction of the down-type fermion (charged lepton or
down-type quark) from the decay of the W boson, both in the W boson
rest frame~\cite{d0_wheli}. The extracted $W$ helicity fractions can be interpreted as
limits on  $f^R_V$, $f^L_T$, or $f^R_T$.
By combining the analysis of
anomalous couplings in single top events with information from the $W$
helicity analysis, posterior probability density distributions
for the anomalous coupling form factors are obtained. This provides
95\% CL limits on anomalous $tWb$ couplings of
$|f_V^R|^2<0.30$,  $|f_T^L|^2<0.05$,  and 
$|f_T^R|^2<0.12$~\cite{anom_combi}.
% anomalous couplings

\section{Top-related Searches for New Physics}
Until today, three generations of quarks and leptons are known in the
SM. A simple extension would be the inclusion of a fourth generation of
fermions. Both collaborations, D0 and CDF, searched for pair
production of  massive fourth
generation quarks, $t^{'}$, assuming the decay into a $W$ boson and a
down-type quark. CDF allows this down-type quark to be $d$, $s$ or
$b$, while D0 assumes a $b$-quark. The search is performed in the
semileptonic final state with at least four jets, of which at least
one has to be an identified $b$-jet at D0. The $t^{'}\bar{t}^{'}$
sample is expected to have a higher fitted top mass and a larger
scalar sum of the lepton and jet $p_T$s, thus the search is performed
as a template fit of these two observables.  Upper limts on
$\sigma(p\bar{p}\rightarrow t^{'}\bar{t}^{'})$ are extracted as
function of the $t^{'}$ mass $m_{t^{'}}$, resulting in lower limits on
$m_{t^{'}}$ at the 95\% CL  of $m_{t^{'}}>285$~GeV at D0~\cite{d0_tprime} using
5.3~fb$^{-1}$ and
$m_{t^{'}}>358$~GeV by CDF~\cite{cdf_tprime} using
5.6~fb$^{-1}$. 

Another top related search recently performed by the CDF collaboration
is a search for massive, collimated jets, which serves as a test of
quantum chromodynamics and can give insights into parton showering
models. The search aims to select events where the decay products of
the top quark are collimated into one single, massive jet. Using
6.0~fb$^{-1}$, CDF requires events with at least one jet cluster with
$p_T>400$~GeV, and high jet masses~\cite{cdf_boostedtop}. The search is performed in the
lepton+jets final state, where high missing transverse energy is
required, and the allhadronic final state, where for each event two jets are required
to have high jet mass and the event has no missing transverse energy. Upper limits
can be set on the $t\bar{t}$ production cross section for two cases.
The resulting  upper limit is $\sigma_{t\bar{t}}<38$~fb at the 95\% CL for events
where at least one top is produced with $p_T>400$~GeV, and
$\sigma_{t\bar{t}}<20$~fb for the pair production of massive objects
produced with $p_T>400$~GeV.

At D0, the possibility of  Lorentz
invariance violation in the top quark sector has been considered, by searching for a time
dependent $t\bar{t}$ production cross section in the lepton plus jets
final state, using 5.3~fb$^{-1}$ of data. Lorentz-violating terms can
be introduced to the SM Lagrangian via an effective field theory in the standard-model extension
(SME)  framework~\cite{liv_theory}. The SME predicts
$\sigma_{t\bar{t}}$ to depend on the sidereal time, due to the change
of the orientation of the D0 detector with the rotation of the Earth
relative to fixed stars. No indication for a time dependent
$\sigma_{t\bar{t}}$ can be observed, and first constraints on Lorentz
invariance violation in the top quark sector are set~\cite{d0_liv}. 
% boosted top
% LIV

\section{Conclusion and Outlook}
In this report, a collection of recent searches for physics beyond the
SM in the top quark sector by the CDF and D0 collaborations has been
discussed. New models have been tested using up to the full Tevatron
data set. No evidence for  physics beyond the
SM has been seen yet.

\section*{Acknowledgements}
I thank my collaborators from CDF and D0
 for their help in preparing the presentation and this
article. I also thank the staffs at Fermilab and
collaborating institutions, and acknowledge the support from STFC.

\end{document}